# The contribution of US broadband infrastructure subsidy and investment programs to GDP using Input-Output modeling

Matthew Sprintson, Edward Oughton

## Abstract

More than one-fifth of the US population does not subscribe to a fixed broadband service despite broadband being a recognized merit good. For example, less than 4% of citizens earning more than US $70k annually do not have broadband, compared to 26% of those earning below US $20k annually. To address this, the federal government has undertaken one of the largest broadband investment programs ever via The Bipartisan Infrastructure Law, with the aim of addressing this disparity and expanding broadband connectivity to all citizens. We examine broadband availability, adoption, and need for each US state, and then construct an Input-Output model to explore the potential structural macroeconomic impacts of broadband spending on Gross Domestic Product (GDP) and supply chain linkages. In terms of macroeconomic impact, the total potential indirect contribution to US GDP by the program could be as high as US $84.8 billion, $32.7 billion, and $9.78 billion for the Broadband Equity, Access, and Deployment program, the Affordable Connectivity Program, and additional programs, respectively. Thus, overall, the broadband allocations could expand US GDP by up to US $127.3 billion (0.10% of annual US GDP over the next five years). Moreover, the broadband packages within the Bipartisan Infrastructure Law could create up to 230,000 jobs (0.14% labor market increase). We contribute one of the first economic impact assessments of the US Bipartisan Infrastructure Law to the literature.

# 1 Introduction

Reliable high-speed broadband is crucial for economic growth and improving productivity. For example, a broadband connection gives firms access to a larger pool of resources, suppliers, and customers, enhancing business growth in both urban and rural regions (DeStefano et al., 2023; Prieger, 2017; Stockinger, 2019) and lowering input prices (LoPiccalo, 2021; Lumpkin & Dess, 2004). Communities desire broadband investments for a variety of reasons, but the economic development benefits are a key motivation, particularly for rural and remote locations that want to boost extra-regional trade (Kumar & Oughton, 2023; Malgouyres et al., 2021; Rodriguez-Crespo et al., 2021). Fledgling companies (as well as established firms) also benefit, as broadband provides entrepreneurs advantages in generating new customers and business opportunities, enabling revenue growth (Chen et al., 2023; Hasbi, 2020; Prieger, 2023; Stephens et al., 2022). Consumers also benefit from good quality broadband infrastructure. Indeed, consumers can access a broader selection of goods and services (Greenstein & McDevitt, 2011), possibly bolstering inter-regional transactions. Investments in broadband can enhance education (Cullinan et al., 2021; Graves et al., 2021; Gu, 2021), expand access to vocational training (Goulas et al., 2021; Rosston & Wallsten, 2020), and develop new ways to participate in the labor force, increasing the productivity of regional and national economies (Gallardo et al., 2021; Mukhalipi, 2018; Pelinescu, 2015).

Given this context, broadband has been a regular news item in current affairs media in recent years across the political spectrum. CNN has reported that broadband infrastructure investment "could make a substantial dent in the country's digital divide" (CNN, 2021), with Fox News stating the plan would "expand broadband access to bring tech jobs to rural America" (Fox, 2020). While decision-makers from different political parties understand that high-speed, reliable broadband connectivity is crucial for societal and economic development, there are often disagreements. At a May 2024 Senate Hearing discussing the Future of Broadband Affordability, Senator John Thune mentioned that the “conditions, rules, and regulations that the administration attached to the program” made it difficult for providers to use the BEAD program, whereas a nonpartisan researcher commented that “these investments may finally put affordable, high-speed Internet within reach of every American in this country.” (*de Wit*, 2024; *The Future of Broadband Affordability*, 2024). Contentions arise with regard to the magnitude of public spending, as well as how broadband infrastructure should be deployed (whether by market methods or government) (Alabama Political Reporter, 2023; Brookings, 2022; Fox, 2023).

Unfortunately, a significant economic divide exists between those with or without a high-speed broadband connection in US communities (Ali, 2022; Valentín-Sívico et al., 2023). Such divisions have become more apparent during the COVID-19 pandemic as rural communities struggled because they could be disproportionately less likely to have a reliable broadband connection (UCSB, 2022). These communities have also had difficulty participating in online commerce (Isley & Low, 2022), accessing essential services, and carrying out transactions online (Lai & Widmar, 2021; Oughton et al., 2024).

In 2020, a bipartisan committee of legislators and policymakers introduced the Infrastructure Investment and Jobs Act, also known as the Bipartisan Infrastructure Law, which includes broadband infrastructure subsidies and resources to boost deployment. The Bipartisan Infrastructure Law consists of three key approaches to expanding broadband coverage and adoption, including (i) the Broadband Equity, Access, and Deployment (BEAD) program, (ii) the Affordable Connectivity Program (ACP), and (iii) smaller programs such as the Tribal Broadband Connectivity Program (TBCP), the Distance Learning, Telemedicine, and Broadband Program (DLTBP), and the Enabling Middle Mile Broadband Infrastructure Program (EMMBIP). Currently, the FCC defines broadband as a connection with a download speed of more than 25 Mbps and an upload speed of more than 3 Mbps (Broadband USA, 2016; FCC, 2022). However, an FCC inquiry recently recommended that the standard be increased to 100 Mbps for download and 20 Mbps for upload (Gorscak, 2023).

The BEAD program received about two-thirds of the US $65 billion the Bipartisan Infrastructure Law allocated towards broadband. With a US $42.45 billion budget, the BEAD program is the largest-ever single congressional allocation towards broadband investment. The BEAD program aims to catalyze broadband infrastructure investment and improve accessibility to a reliable, fast Internet connection to qualifying US citizens. The budget was split between the states and several territories through a formula and prioritized building wireline broadband for unserved locations in high-cost areas. By working with broadband providers, states and territories aim to construct or improve broadband infrastructure in the United States (Congressional Research Service, 2023). In addition to infrastructure challenges, barriers ranging from adoption to technology and implementation costs need to be overcome (Canfield et al., 2019).

The Affordable Connectivity Program, administered by the Federal Communications Commission (FCC), allocates a subsidy for households to purchase broadband connections. Eligible families can receive a discount of up to US $30 per month, while those on tribal lands can receive up to US $75 per month. The program allocates US $14.2 billion for broadband investment and provides up to a US $100 discount for a computer or tablet (*Affordable Connectivity Program*, 2023). This program aims to make broadband more accessible to consumers and thus increase the demand for broadband, as decreasing broadband prices have been shown to be positively associated with increased broadband penetration (Abrardi & Cambini, 2019; Flamm & Chaudhuri, 2007; Rosston & Wallsten, 2020).

Finally, three smaller programs are considered within the scope of this paper. The Tribal Broadband Connectivity Program focuses on broadband deployment on tribal lands, which comprise approximately 2.3% of US land area, with the Bipartisan Infrastructure Law allocating $2 billion to invest in servicing these communities (Vincent et al., 2017). The TBCP targets increasing access in Tribal areas through awards, with one recent example being the allocation of $19.8 million to the Chippewa Tribe living in Boise Forte to install fiber cables in their community (Oxendine, 2023). Grants for native territories have been found to positively affect broadband penetration (Korostelina & Barrett, 2023; Pipa et al., 2023).

The DLTBP, administered by the US Department of Agriculture, was created with a US $2 billion budget to provide grants to local governments, non-profit organizations, and businesses to expand access to telecommunications infrastructure for distance learning and telehealth services. The EMMBIP, administered by the NTIA, was a US $1 billion investment into middle-mile infrastructure, broadband infrastructure that does not come in direct contact with end-users. This project hopes to extend current broadband networks and make high-speed Internet networks more reliable and less expensive for underserved areas.

With this context in mind, this paper subsequently analyzes the economic impact of these three broadband investment and subsidy programs. This is undertaken by first exploring the availability, adoption, and need for broadband on a state-by-state level. Then secondly, developing an Input-Output (IO) model to quantify the structural macroeconomic effects in GDP and employment terms. Finally, the potential supply chain linkage effects are quantified for different industrial sectors. The research questions include the following:

1. To what extent does the Bipartisan Infrastructure Law allocate funding to unconnected communities in need?
2. What are the GDP and employment impacts of the three funding programs within the Bipartisan Infrastructure Law?
3. How are supply chain linkages affected by allocations from the Bipartisan Infrastructure Law?

In Section 2, a literature review is carried out examining studies pertaining to the research questions, before a method is presented in Section 3. Modeling results will be presented in Section 4, and we then return to the research questions in Section 5 to discuss the ramifications of the findings in the broader policy context. Finally, the research conclusions, contributions to the literature, and limitations of the approach will be presented in Section 6.

# 2 Literature Review

## 2.1 Reviewing Broadband Infrastructure's Impact on the Economy

Access to broadband has many benefits, as examined in this review. While broadband is a necessary but not sufficient factor for development in modern economies, connecting more people to a faster Internet removes a significant barrier that constrains many communities, especially those rural and remote. However, it is also possible that broadband deployment could have a negative impact on employment in some industrial sectors (Zhou et al., 2022). In industries where Internet access cannot replace less efficient labor activities, such as manufacturing, increased broadband adoption positively affects productivity and output (Jung & López-Bazo, 2020; Zhang et al., 2022). However, the return from broadband investment generally depends on the wider level of availability and adoption within an economy. For example, regions with poorer broadband infrastructure generally see a greater contribution to employment and economic growth when finally deployed (Pradhan et al., 2018; Shideler & Badasyan, 2007). However, for regions that already have significantly comprehensive broadband infrastructure, there are generally diminishing returns to scale, with increased infrastructure spending leading to diminishing returns on investment (as is common in other infrastructure sectors such as transportation, energy, etc.).

In terms of macroeconomic impacts, the deployment of broadband infrastructure is generally found to have a positive effect on economic growth (Koutroumpis, 2009). Specifically, increased broadband penetration leads to positive impacts on GDP after more than half of the population has gained access to the Internet. For example, a 1% increase in broadband penetration yields a 0.02% increase in GDP in areas with low broadband penetration and 0.03% elsewhere. In contrast, one assessment estimates broadband's marginal effect on GDP using pricing models and "willingness-to-pay" estimates, finding that approximately US $8.3-$10.6 billion was generated from US broadband investment prior to 2006. The study's conclusions suggest that broadband growth results directly in economic benefits from improvements in public health, education, local growth, and employment (Greenstein & McDevitt, 2011).

Moreover, another approach utilizing a Cobb-Douglas production function alongside a generalized least squares model to obtain regression specifications quantifies the impacts of broadband on total output (Ghazy et al., 2022). The analysis finds that broadband has a significant positive association with economic output at the 1% significance level. Furthermore, the regression also estimates a 1% increase in connectivity leads to a 0.63% increase in new business entry into the market, concluding that a secure, dependable broadband connection provides entrepreneurs more incentive to enter a market, as increases in broadband penetration lead to an improved business development environment.

Broadband infrastructure investment expands opportunities for entrepreneurs and encourages business creation (Deller et al., 2022; Luo et al., 2022; Stephens et al., 2022). For example, improvements in broadband infrastructure are found to have enhanced firm formation throughout a wide variety of business sectors (Duvivier et al., 2021). This is for two key reasons. Firstly, broadband access encourages entrepreneurs to enter the market. Secondly, broadband can relieve positional barriers that discourage entry into the market. Therefore, when regressed against entrepreneurial measures, like project investment, broadband infrastructure investment indicates a significantly positive effect with a coefficient of 0.03. The analysis also suggests there are benefits to innovation from broadband investment (Han et al., 2023; Rampersad & Troshani, 2020; Xu et al., 2019), with an additional 1% penetration associated with an increase of up to 1.4% in filed patents (analogous to innovation) (Yang et al., 2022).

To conclude, there is strong evidence that investing in broadband infrastructure positively affects a range of economic metrics, including industrial output, productivity, entrepreneurship, employment and innovation.

## 2.2 Previous Research into IO Modeling of Broadband Investment

Input-Output (IO) Modeling is a macroeconomic method developed by Wassily Leontief and awarded the Nobel Prize in Economic Sciences in 1973. The approach represents economies as a matrix where the rows represent sectoral outputs and the columns represent sectoral inputs. Assuming constant economic returns, one splits an economy into $n$ sectors to analyze intersectional demand. The final economic output is represented in the dollar value of the goods and services that a sector produces. The value is calculated by summing the amount purchased as an input by other sectors of the economy and the amount purchased as a final good by consumers. When incorporating imports, government spending, value-added analysis, and other factors into separate rows and columns, it is possible to use this matrix to analyze the total output of an economy. When adjusting the inputs for each sector, it is possible to quantify the marginal effects of policy and investment either for a specific industry or the wider macroeconomy (Leontief, 1986).

Government investment in broadband, Internet, and mobile networking infrastructure has been shown to have a range of positive economic growth impacts through GDP, labor, and other heuristics (Abrardi & Cambini, 2019; Kim et al., 2021). Infrastructure investment impacts the economy through discrete sectoral spillovers (Schreiner & Madlener, 2022; Välilä, 2020). A quantifiable metric for identifying these spillovers is the analysis of the sectoral impacts on each other (Jimmy & Falianty, 2021). For example, the economic consequences of infrastructure development can be considered through the lens of Macroeconomic Growth

Theory (Carlsson et al., 2013). Investment is represented within these models as structural changes to those sectoral effects by amplifying some sectors and damping others (Nieto et al., 2020; Sievers et al., 2019).

The Leontief approach has been used to quantitatively analyze broadband investment appropriations in the American Recovery and Reinvestment Act of 2009. A measure of proportional growth can be captured for the interactions between industrial sectors resulting from additional investment. This enables direct, indirect, and induced effects to be quantified for the total output of each sector. One study found that a 1% in broadband availability led to a 0.03% increase of employment in metropolitan counties. The study also finds that broadband tends to lead to significant job creation in service industries and discusses a study that reported the employment impact of broadband deployment to 95% of homes could be up to 140,000 jobs (R. Katz & Suter, 2009). Another study finds the spillover effects of infrastructure projects and uses them to explore future economic outputs (Dimitriou et al., 2015). For example, business services were found to have a growth multiplier of 2.2, while post and telecommunications would potentially see a growth multiplier of 1.55. This figure denotes the growth in the telecommunications sector for each dollar invested in project construction. Additionally, one assessment uses an IO model to evaluate German broadband investment, finding that an investment of 36 billion (euros) could increase GDP by approximately 171 billion (euros) (Katz et al., 2010).

Both Leontief demand-side IO modeling (discussed mathematically in Section 3.1) and Ghosh supply-side IO modeling (discussed mathematically in Section 3.2) have been used to study economic structures and the macroeconomic effect of stimulus. While Leontief IO Models study the upstream effects of downstream stimuli, Ghosh IO models study the downstream effects of upstream stimuli (Miller & Blair, 2009). Ghosh modeling has been used to study the downstream impacts of Information and Communication Technology and natural hazards by investigating forward linkages from impacted sectors of the economy. By augmenting value-added as a result of a stimulus, the researchers analyzed the downstream consequences through a supply-side model. In both cases, Ghosh supply-side modeling was a valuable counterpart to Leontief modeling and employed Static IO methodology to study downstream impacts (Oughton et al., 2017; Yuan et al., 2024).

Static IO modeling is one of many techniques researchers employ to study the impact of broadband, telecommunications, or other information infrastructure. Other contemporary research has employed production functions (Appiah-Otoo & Song, 2021), CGE modeling (Hwang & Shin, 2017), or other steady-state models (de Clercq et al., 2023). IO Modeling is best suited for providing insights into economic structures expressed through a system of linear equations (Zheng et al., 2023). While CGE modeling produces a long-run lower-bound estimate focusing on behavioral effects like substitution, recovery, or other long-term relationships, IO modeling provides a short-run upper-bound analysis of inter-sectoral linkages and macroeconomic effects (Macmillan et al., 2023).

Previous research has justified the use of IO modeling when analyzing the macroeconomic effects of investment alongside economic structures like forward linkages, backward linkages, and the intersectoral supply chain (Akbari et al., 2023; Sahani et al., 2023). Contemporary studies have used IO modeling architecture to conduct similar structural analyses for investments into Information and Communication Technologies and other infrastructure systems (Choe et al., 2023; Stamopoulos et al., 2022; Vu & Nguyen, 2024; Zhang et al., 2022). For example, investment impacts on regional economic sectors are estimated using an IO model for US energy infrastructure investment via supply and use tables from the Bureau of Economic Analysis and the IELab US Multi-Regional IO database (Faturay et al., 2020). Another study uses IO modeling to assess port infrastructure on other sectors of the Chinese economy. The methodology allowed the researchers to separate economic effects between industries, showing that the construction, chemical, and transportation sectors would be affected the most by port closures, with overall GDP reducing as much as US $90 billion (Wang & Wang, 2019).

These sources illustrate that IO modeling is an established approach for evaluating the macroeconomic impacts of infrastructure investment, especially when considering inter-industry linkages.

## 2.3 Context of the Bipartisan Infrastructure Act through Previous Research

Policy choices can help maneuver broadband networks to achieve more positive outcomes (Oughton et al., 2023; Oughton, 2023), but efforts to close broadband gaps have not yet managed to do so (King & Gonzales, 2023). Currently, only 77% of people in the United States subscribe to a broadband connection (Pew Research Center, 2021). Many unconnected Americans lack access because of connection prices, insufficient technology, scarce information, and inadequate government policies (Bauer, 2023; Oughton et al., 2022; Rosston & Wallsten, 2020). A significant divide exists between urbanization and income levels in US households (Rothschild, 2019). For example, with 82% of urban households having a fixed broadband connection, only about 70% of rural areas have the same connection available (Li et al., 2023; United States Census Bureau, 2021).

Most Native Americans living in tribal areas do not have access to the same broadband infrastructure that other communities have. There are numerous reasons - progress is hampered by a lack of trust, limited resources, and insufficient education (Korostelina & Barrett, 2023). Government investment could lead to increases in broadband connectivity, but substantial disparities still prevail (Andres et al., 2024; Healy et al., 2022). For example, 55.6% of Native American tracts have seen an expansion in broadband providers between 2004 and 2014, with 27.7% of Native American tracts now having an above-average number of providers (Mack et al., 2022). However, the percentage of Native households with Internet access is still 21% lower than in the surrounding areas (Bauer et al., 2022). For the US government to ensure everyone can participate in a modern online economy, there needs to be more infrastructure built to support tribal broadband and investment in education and awareness (Duarte et al., 2021; Hudson et al., 202; Mack et al., 2023).

As found in previous studies, Figure 1 emphasizes empirically the association between broadband inequality and income level (Deng et al., 2023; Houngbonon & Liang, 2020; Wolfson et al., 2017). This problem is exacerbated as broadband adoption has been found to increase wages and hiring in a community (Poliquin, 2021; Yin & Choi, 2023). Thus, communities with reliable broadband connectivity often have higher earnings and more opportunities, aggravating the inequality shown in the figure (Consoli et al., 2023; Mathews & Ali, 2023). Figure 1 (A) shows that, on average, communities with a mean income of less than $20,000 have poorer access to broadband than wealthier communities. This can constrain economic opportunities in poorer areas, meaning citizens in these areas are disadvantaged in seeking new employment options and adapting to changing labor markets. Figure 1 (B) shows us that the disparities in connectivity,

separated by income, are present when we compare regions with larger broadband allocations. 69.3% of those making under US $20k in states receiving over US $1,350 have access, whereas 75.1% have access in states receiving less than US $50k.

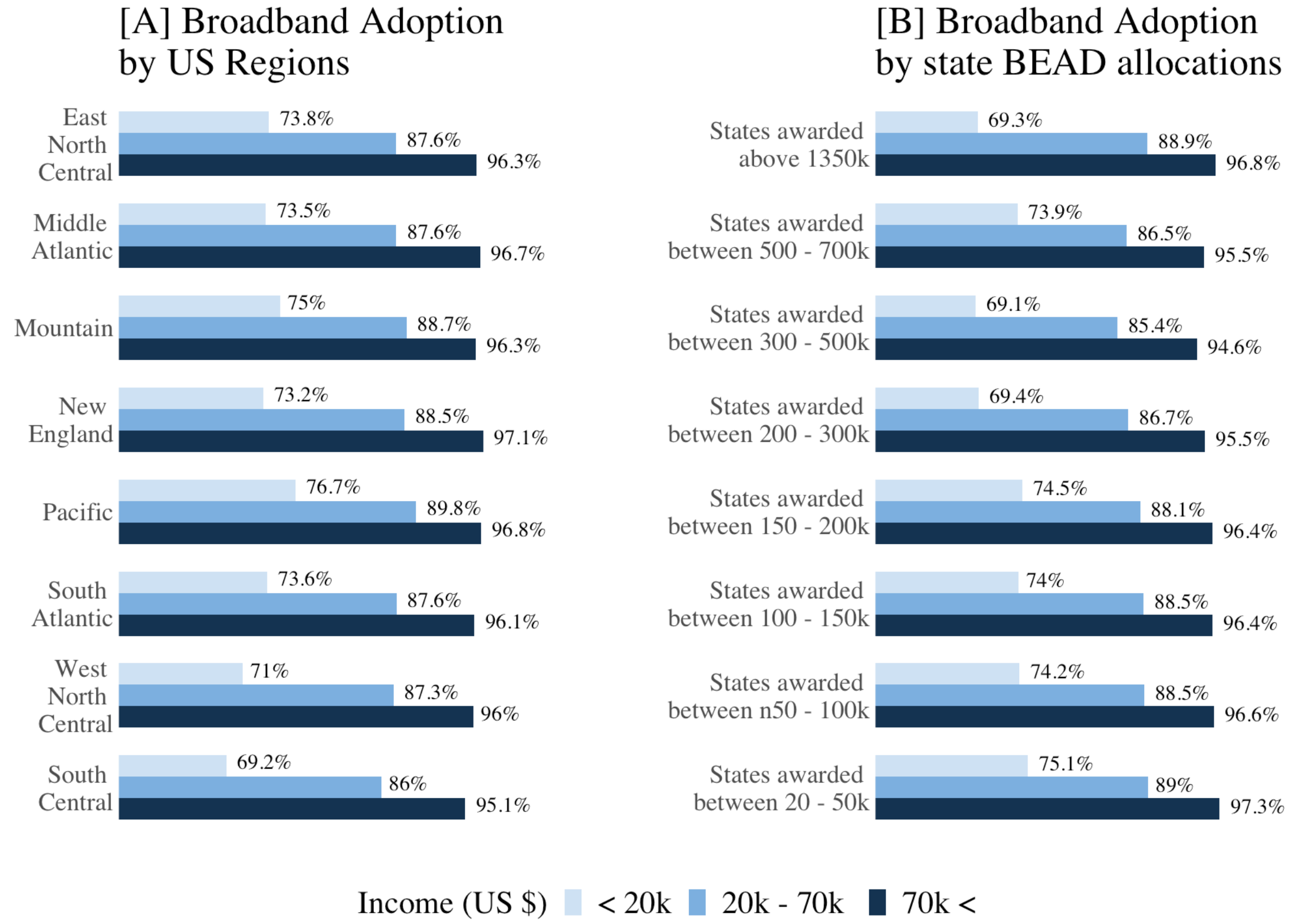

Figure 1: Location and income are key determinants of broadband adoption
(A),(B): U.S. Census Bureau, 2021 American Community Survey (UCSB, 2022)

Finally, 20.9% of tribal lands still lack access to reliable broadband (Hutto & Wheeler, 2023). The adverse effects of the lack of dependable broadband were shown throughout the COVID-19 pandemic when many tribal communities were restricted from accessing schools and working opportunities because they did not have a reliable connection to broadband infrastructure (Kroll, 2023; Le-Morawa et al., 2023; Levin et al., 2023). Broadband connectivity can be efficiently implemented through inter-tribal communication throughout the community and community-specific broadband dissemination (Gellman et al., 2021). The Tribal Connectivity Program's awards aim to support the deployment of required broadband infrastructure by working in tandem with affected communities (*Tribal Broadband Connectivity Program*, 2023).

Now that a thorough literature review has been undertaken, the methods for analysis will be presented.

# 3 Methods

In this Section, we define terms associated with our analysis, mathematically describe various methods for assessing the macroeconomic contribution of infrastructure to the economy, and discuss how we will use data to find estimates in the context of the research questions.

Throughout this paper, we use the notation defined by the seminal work Miller and Blair (2009). A "direct impact" denotes the necessary direct outputs to satisfy the increase in final demand. If the federal government spends $x on broadband, then the direct impact into the sector is commensurately $x. An "indirect effect" is the excess output stimulated by the spending. In other words, it represents the ripple effect throughout the economy. These effects can be broken down into "upstream" and "downstream" effects. When a sector increases production, it demands more input goods from other sectors; this relationship is categorized as an "upstream effect" and can be modeled using a Leontief matrix. However, when a sector increases production, it means more supply is available for other sectors to use as inputs and can influence output in associated sectors - this relationship is categorized as a "downstream effect" and can be modeled using a Ghosh matrix. A "Type I Multiplier" refers to the relationship between the investment into a sector and the direct plus indirect effects of that investment. Furthermore, a "Keynesian Multiplier" is defined as the GDP change per $1 of direct investment. It represents the relationship between investment into a sector and the direct plus indirect and induced effects of that stimulus.

We continue by articulating a general macroeconomic system model based on IO before detailing different demand-side and supply-side evaluation methods for broadband infrastructure.

## 3.1 Leontief Input-Output (IO) Modeling

The IO approach begins by dividing the economy into $n$ sectors, such as agriculture, manufacturing, energy, telecommunications, and other such industries. The model denotes inter-sectoral transfers as $z_{ij}$, representing the dollar value of transfers from sector $i$ to sector $j$. For example, if sector $1$ is electricity and sector $2$ is telecommunications, $z_{12}$ represents the dollar value of electricity sold to the telecommunications industry as an input. The other metric is $f_i$, which denotes the final demand for the good or service from sector $i$. The final demand includes purchases by consumers, investments from businesses, purchases from the government, and net exports.

The final output of sector $i$ is denoted as $x_i$, measured here in dollars, and represents the sum of all monetary values of the transactions from other inter-industry sectors, as well as the final demand for goods or services, stated in equation (1) as follows.

$$x_i = \sum_{j=1}^{n}(z_{ij}) + f_i \tag{1}$$

Thus, we can list the linear equations of all $n$ industries in rows, as detailed in equation (2).

$$\begin{gathered} x_1 = z_{11} + \cdots + z_{1j} + \cdots + z_{1n} + f_1 \\ \vdots \\ x_i = z_{i1} + \cdots + z_{ij} + \cdots + z_{in} + f_i \\ \vdots \\ x_n = z_{n1} + \cdots + z_{nj} + \cdots + z_{nn} + f_n \end{gathered} \tag{2}$$

Moreover, the vector for the final output is $x$, as outlined in equation (3).

$$x = \begin{bmatrix} x_1 \\ \vdots \\ x_i \\ \vdots \\ x_n \end{bmatrix} \tag{3}$$

Likewise, the matrix of inter-industry transactions and final demand can be represented as $Z$ and $F$, respectively, as we present in equations (4) and (5).

$$Z = \begin{bmatrix} z_{11} + \cdots + z_{1j} + \cdots + z_{1n} \\ \vdots \\ z_{i1} + \cdots + z_{ij} + \cdots + z_{in} \\ \vdots \\ z_{n1} + \cdots + z_{nj} + \cdots + z_{nn} \end{bmatrix} \tag{4}$$

$$F = \begin{bmatrix} f_1 \\ \vdots \\ f_i \\ \vdots \\ f_n \end{bmatrix} \tag{5}$$

As each sector of the economy can be modeled in a row of the matrices above, the total expression for the total $n$ industries in the economy can be modeled in a system of linear equations represented in equation (6).

$$x = Zi + F \tag{6}$$

Where $i$ represents an identity matrix of dimensions $n$ by $n$.

Each row of the matrix Z represents the aggregate output of its respective sector. To measure spillovers across industries, we can analyze technical coefficients $a_{ij} = z_{ij} / x_j$. For two industries $i$ and $j$, the technical coefficient $a_{ij}$ is equal to the dollar's worth of input from $i$ to the output of sector $j$.

Because $z_{ij} = a_{ij} * x_j$, we can replace the respective coefficients algebraically, as detailed in equation (7).

$$\begin{aligned} x_1 &= a_{11}x_1 + \cdots + a_{1j}x_j + \cdots + a_{1n}x_n + f_1 \\ &\vdots \\ x_i &= a_{i1}x_1 + \cdots + a_{ij}x_j + \cdots + a_{in}x_n + f_i \\ &\vdots \\ x_n &= a_{n1}x_1 + \cdots + a_{nj}x_j + \cdots + a_{nn}x_n + f_n \end{aligned} \tag{7}$$

Distributing, as per equation (8).

$$
\begin{gathered}
x_1 - a_{11}x_1 - \cdots - a_{1j}x_j - \cdots - a_{1n}x_n = f_1 \\
\vdots \\
x_i - a_{i1}x_1 - \cdots - a_{ij}x_i - \cdots - a_{in}x_n = f_i \\
\vdots \\
x_n - a_{n1}x_1 - \cdots - a_{nj}x_i - \cdots - a_{nn}x_n = f_n
\end{gathered} \tag{8}
$$

Factoring, as per equation (9).

$$
\begin{gathered}
(1 - a_{11})x_1 - \cdots - a_{1j}x_j - \cdots - a_{1n}x_n = f_1 \\
\vdots \\
-a_{i1}x_1 - \cdots + (1 - a_{ij})x_i - \cdots - a_{in}x_n = f_i \\
\vdots \\
-a_{n1}x_1 - \cdots - a_{nj}x_j - \cdots + (1 - a_{nn})x_n = f_n
\end{gathered} \tag{9}
$$

Thus, as per the system of linear equations (10).

$$(I - A)x = F \tag{10}$$

Where *A* is the n by n matrix of technical coefficients as represented below, in equation (11).

$$
A = \begin{bmatrix}
a_{11}, \cdots a_{1j}, \cdots a_{1n} \\
\vdots \\
a_{i1}, \cdots a_{ij}, \cdots a_{in} \\
\vdots \\
a_{n1}, \cdots a_{nj}, \cdots a_{nn}
\end{bmatrix} \tag{11}
$$

And the inverse of *(I-A)* is denoted as *L*, the Leontief-inverse matrix, as per equation (12).

$$x = LF \tag{12}$$

Therefore, we can use this formula to quantify the output generated by a shift in final demand for individual industrial sectors, or the whole economy.

## 3.2 Ghosh Supply Side Assessment Methods for Infrastructure

The Ghosh Supply Side model, as described by Miller and Blair (2009), focuses on the downstream impacts of economic stimulus. While the Leontief modeling method, described in Section 3.1, estimates the upstream impacts due to changes in final demand, the Ghosh modeling method estimates the downstream impact of changes in value-added on the supply-side. Therefore, we use the Leontief approach when analyzing programs that stimulate demand and the Ghosh approach when analyzing programs that stimulate supply.

The Ghosh Supply Side model measures the changes in availability of inputs on industrial output. While the Leontief matrix relies on technical coefficients $a_{ij}$, the Ghosh matrix relies on allocation coefficients $b_{ij} = z_{ij} / x_i$. The allocation coefficient measures the value of transactions from sectors $i$ to $j$ divided by the output of sector $i$.

The allocation coefficients can be made into a matrix $B$ similar to $A$, as detailed in equation (13).

$$\begin{bmatrix} \frac{1}{x_1} & 0 & 0 & 0 & 0 \\ 0 & \ddots & 0 & 0 & 0 \\ 0 & 0 & \frac{1}{x_i} & 0 & 0 \\ 0 & 0 & 0 & \ddots & 0 \\ 0 & 0 & 0 & 0 & \frac{1}{x_n} \end{bmatrix} \begin{bmatrix} z_{11} & \cdots & z_{1j} & \cdots & z_{1n} \\ \vdots & & \vdots & & \vdots \\ z_{i1} & \cdots & z_{ij} & \cdots & z_{in} \\ \vdots & & \vdots & & \vdots \\ z_{n1} & \cdots & z_{nj} & \cdots & z_{nn} \end{bmatrix} = \begin{bmatrix} \frac{z_{11}}{x_1} & \cdots & \frac{z_{1j}}{x_1} & \cdots & \frac{z_{1n}}{x_1} \\ \vdots & & \vdots & & \vdots \\ \frac{z_{i1}}{x_i} & \cdots & \frac{z_{ij}}{x_i} & \cdots & \frac{z_{in}}{x_i} \\ \vdots & & \vdots & & \vdots \\ \frac{z_{n1}}{x_n} & \cdots & \frac{z_{nj}}{x_n} & \cdots & \frac{z_{nn}}{x_n} \end{bmatrix} = \begin{bmatrix} b_{11} & \cdots & b_{1j} & \cdots & b_{1n} \\ \vdots & & \vdots & & \vdots \\ b_{i1} & \cdots & b_{ij} & \cdots & b_{in} \\ \vdots & & \vdots & & \vdots \\ b_{n1} & \cdots & b_{nj} & \cdots & b_{nn} \end{bmatrix} = B \tag{13}$$

Thus, as per equation (14).

$$\begin{aligned} \hat{x}^{-1} Z &= B \\ Z &= \hat{x} B \end{aligned} \tag{14}$$

From the previous Section, we know that Ax = Z, as in equation (15).

$$\begin{aligned} Ax &= \hat{x} B \\ \hat{x}^{-1} Ax &= B \end{aligned} \tag{15}$$

The summation of the inputs of an industry plus its value added is equal to its output, per the IO table.

Thus, as per equation (16).

$$x_j = \sum_{i=1}^{n} (z_{ij}) + v_j \tag{16}$$

Similarly, as per equation (17).

$$\begin{aligned} x' &= i'Z + v' \\ x' &= i'xB + v' \\ x' - x'B &= v' \\ x'(I - B) &= v' \\ x' &= v'(I - B)^{-1} \\ x' &= v'G \end{aligned} \tag{17}$$

The Ghosh matrix, *G*, is equal to $(I\text{-}B)^{-1}$ and allows us to see the changes in final output from the changes in value added.

## 3.3 Data and Application

We will source IO tables from the Bureau of Economic Analysis (*Input-Output Data Tables*, 2023). The sector for broadband spending is Telecommunications (NAICS 517) (US Census Bureau, 2023). For our augmented final demand/value-added vectors, we can isolate the investments and introduce them into their respective sectors by entering 0 for all other sectors. We can calculate broadband's final demand/value-added by summing the government's subsidies and the increased household spending on broadband connections because they can now afford those services. As discussed in previous literature, we estimate the job creation of the investment package by creating a linearized matrix representation for each job based on the GDP output of each sector. We then augment the matrix with the estimated total GDP growth for each sector and analyze the resultant job growth matrix within the context of the Bipartisan Infrastructure Law.

| Program | Agency | Purpose | Budget ($Bn) | Household contribution ($Bn) | Model Employed |
|---|---|---|---|---|---|
| BEAD | NTIA | To expand broadband equity and access and to deploy broadband service to unserviced homes | 42.45 | 0 | Ghosh |
| ACP | FCC | To provide a subsidy ($30+) for connected homes to buy broadband service | 14.2 | $3.05 \pm 0.21$ | Leontief |
| TBCP | NTIA | To deploy broadband to tribal regions without broadband access | 2 | 0 | Ghosh |
| DLTBP | USDA | To provide grants to help rural communities access broadband facilities for distance learning or telemedicine | 2 | 0 | Leontief |
| EMMBIP | NTIA | To make direct investments into broadband supply infrastructure that does not connect to an end-user | 1 | 0 | Ghosh |

Table 1: The constituent programs of the Bipartisan Infrastructure Law's funding and modeling method. (Infrastructure Investment and Jobs Act, 2021)

As seen in *Table 1*, our analysis includes the BEAD program (Page 2470), ACP (Page 2541), TBCP (Page 2471), DLTBP (Page 2465), and the EMMBIP (Page 2157) because they include direct investments into broadband infrastructure or subsidies. Our analysis notably excludes the Digital Equity Act Programs ($2.75 Bn, Defined Page 2090, Allocated Page 2471) as the program is an investment into education. While the literature review notes that education is critical for closing the digital divide, this program is not a structural investment, and Input-Output modeling would not capture its macroeconomic effect. Therefore, we consider it exogenous to the model (Infrastructure Investment and Jobs Act, 2021).

According to the ACP, US $14.2 billion is available in subsidies (FCC, 2023). As these are awarded as subsidies, each allocation leads to some household spending, which also contributes to the final demand. In total, if each subsidy is about US $30 per month, or US $360 a year, there will be 39.4 million household recipients throughout the program, assuming we treat this as though households must re-enroll each year. The ACP survey, conducted by the FCC in 2023, reported that approximately 20 to 23% of subsidy recipients were new subscribers (*ACP Consumer Survey*, 2023). Thus, approximately 20-23% of these 39.4 million households will contribute about US $30 themselves monthly, as the average price of broadband is approximately US $61 (*Broadband Affordability*, 2023; Shevik, 2023; Wilson, 2023), minus the subsidy, which is about US $360 a year. We assume that the induced spending of other subscribers is exogenous to the model, as they likely used the extra $30 per month elsewhere or used it to upgrade their subscription. So, households will provide an additional US $2.84-3.26 billion in spending, and the final demand increase for the ACP will be US $29.1 billion.

Ghosh modeling architecture is appropriate to estimate the possible indirect effects of the BEAD program, the TBCP, and the EMMBIP, as they are associated with direct investments into broadband infrastructure. Because these programs seek to enlarge, ensure, or strengthen the supply of broadband, their downstream effects can be studied through the Ghosh method described in Section 3.2. The upstream impacts of the ACP and DLTBP can be modeled with the Leontief IO method, as these programs would increase the final demand for broadband. Therefore, these programs can be modeled with the Leontief IO method, as described in subsection 3.1 (*Mathematical Derivation of the Total Requirements Tables for Input-Output Analysis*, 2017).

Using this methodology, we can also produce, analyze, and contextualize the multipliers for each program. Now that our methods are established, we can examine the results of our study.

# 4 Results

The results are reported for the three research questions defined in the introduction. In Section 4.1, we will discuss which states receive broadband investment, in Section 4.2, we will report the GDP estimates of our models, and in Section 4.3, we will examine the supply-chain linkages associated with the telecommunications sector.

## 4.1 To what extent does the Bipartisan Infrastructure Law allocate funding to unconnected communities in need?

Research on broadband infrastructure and investment has struggled because of data uncertainty, data collection limitations, and unreliable assumptions of heterogeneous coverage and accessibility throughout states, coverage areas, and socioeconomic groups (Lobo et al., 2020; Mack, 2019). The following graphs present aggregated state-level data that has limitations due to the assumption of heterogeneity of broadband across space, socioeconomic groups, and regional contexts. These limitations and their implications, including the human factor in broadband adoption and access, are discussed later in the paper.

Figure 2 (A) indicates that the states with the highest total BEAD allocation are Texas ($3.31 Bn, 7.80%), California ($1.86 Bn, 4.39%), Missouri ($1.74 Bn, 4.09%), Michigan ($1.56 Bn, 3.67%), and North Carolina ($1.53 Bn, 3.61%). The states with the least include the District of Columbia ($.101 Bn, .24%), Delaware ($.108 Bn, .25%), Rhode Island ($.108 Bn, .26%), and North Dakota ($.130 Bn, .31%).

Figure 2 (B) presents the BEAD allocation using the number of unconnected households as the denominator. Alaska is an outlier, spending US $38,802 per unconnected household (729% higher than the mean per household). Now, Texas and California rank towards the lower end of the spectrum, spending US $2,900 and US $1,796 (38.1% and 61.6% lower than the mean per household). States with a larger rural population, like West Virginia (51.3% Rural), Montana (44.1% rural), and Wyoming (33.2% rural), rank higher now, as they are spending on average US $11,960 per home. Only Alabama (41.0% rural) and Missouri (29.6% rural) rank high by both absolute and relative measures of investment.

ACP Household Enrollment is highly correlated with states that have higher populations, as shown in Figure 2 (C), seeing as states like California (2.30 mn), Texas (1.49 mn), Florida (1.39 mn), New York (1.43 mn), and Ohio (1.00 mn) are those with higher enrollment. However, when we measure enrollment using total households as the denominator in Figure 2 (D), we see a different ranking. High-enrollment states show relatively average enrollment per household, like California (2.30 mn, 17.1%), Texas (1.49 mn, 13.8%), and New York (1.43 mn, 18,7%). However, states across the South have relatively high enrollment rates potentially driven by lower incomes and higher per-capita poverty rates. This trend is exemplified in states like Mississippi (.218 mn, 19.3%), Alabama (.366 mn, 18.6%), and an extreme outlier, Louisiana (.474 mn, 26.6%). The Rust-belt region also has high enrollment, including states like Ohio (1.00 mn, 20.7%) and Kentucky (.404 mn, 22.6%).

Furthermore, states like Ohio (20.7%), New Mexico (20.2%), and Kentucky (22.6%) have greater enrollment, which is also a potential indicator of improved broadband access (comparatively with the pre-ACP period). Households without access to broadband providers gain no additional utility in enrolling in a subsidy program for broadband.

Many researchers have noted the significant human factor involved with broadband availability, enrollment, and eventual adoption. Human capital, certain occupations, availability, and enrollment are critical factors for broadband to have an economic impact in communities that investments target (Mack & Faggian, 2013; Whitacre & Brooks, 2014). Moreover, research has shown that individual state policy has a considerable impact on regional broadband availability, which suggests that higher ACP enrollment is related to ground-level coordination efforts and policy communication greater than a lack of accessibility (Whitacre et al., 2013; Whitacre & Gallardo, 2020).

Other steps are critical to note, like adoption, availability, firm behavior, socioeconomic groups, and income (Gallardo & Whitacre, 2024; Whitacre et al., 2014). However, when considering the graphs in Figure 2 at face value, ACP enrollment can serve as a marker for unaffordable broadband availability, as participants enroll when they cannot afford service that is already available. Additionally, BEAD grants could serve as a marker for lack of broadband availability, as the allocation formula incorporated unserviced locations and high-cost unserviced locations to calculate the grants (Pew Research Center, 2023).

Therefore, states with high BEAD allocations but low ACP enrollment could be receiving funds to deploy broadband because communities in-need either i) lack communication that they can subscribe to ACP to afford existing infrastructure or ii) are unable to subscribe to ACP because they lack access to reliable high-speed Internet infrastructure. This includes states like Montana ($11,900 per unenrolled household, 10.7% ACP enrollment), Wyoming ($13,800 per unenrolled household, 7.63% ACP enrollment), and West Virginia ($10,300 per unenrolled household, 15.5% enrollment). Additionally, there are states with the opposite characteristics, like California ($1,800 per household, 17% ACP enrollment) and Ohio ($1,400 per household, 20.7% enrollment), which could have lower BEAD allocations because there are already considerable broadband networks and the largest barrier to entry is affordability.

Thus, with respect to other social, infrastructural, and economic considerations, low BEAD allocations associated with high ACP enrollment could suggest unaffordable connections are present for communities in need. Additionally, high BEAD allocations associated with low ACP enrollment could suggest that many unconnected households lack service and investments focus on increasing availability.

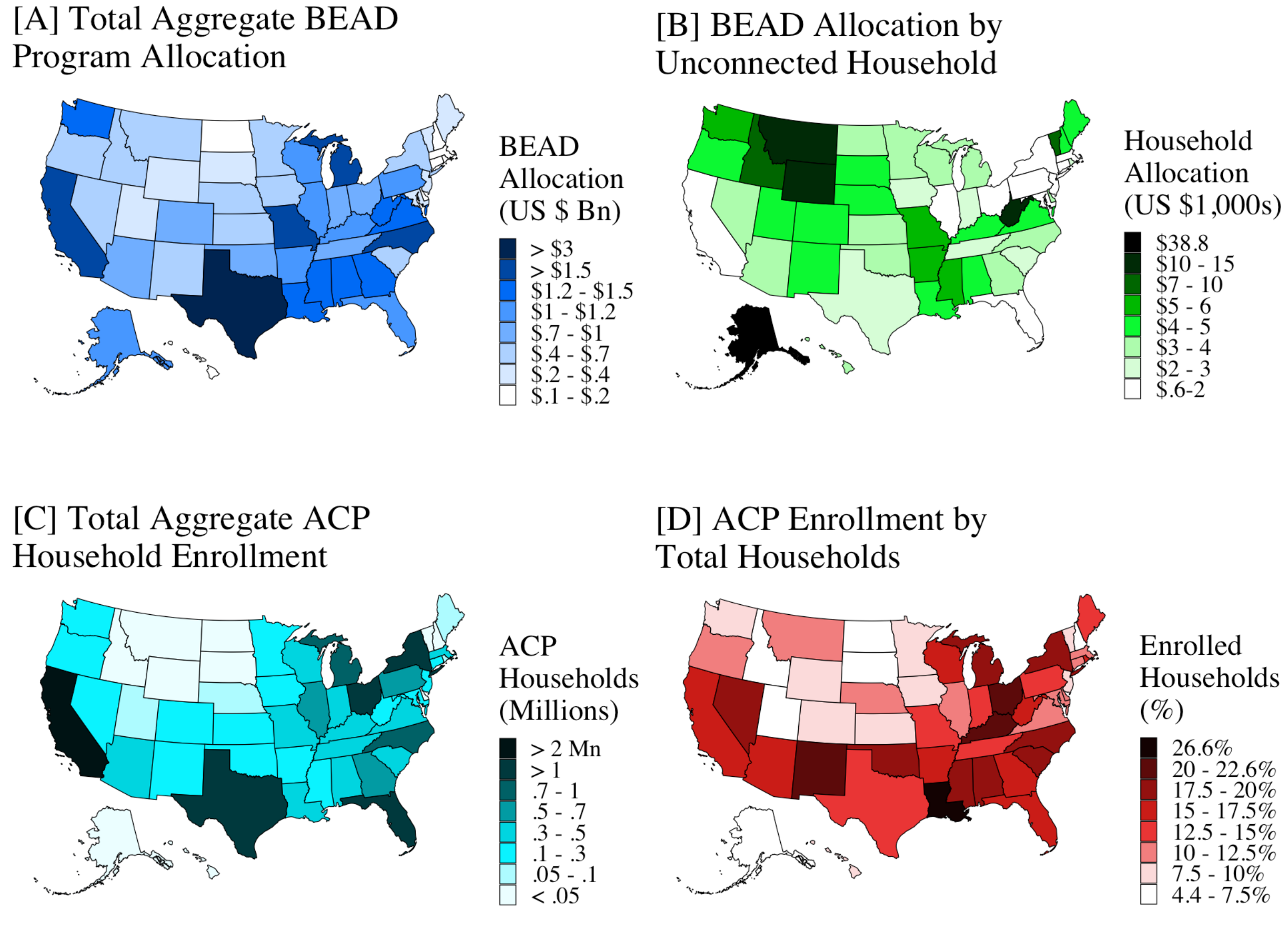

Figure 2: Visualization of absolute and relative BEAD and ACP allocations by state

## 4.2 What are the GDP and employment impacts of the three funding programs within the Bipartisan Infrastructure Law?

Via the model we specify, we estimate that the total potential macroeconomic impact of the US $61.7 billion direct federal investment allocated through the selected programs of the Bipartisan Infrastructure Law could result in a total US $127.3 Bn increase in GDP (subject to the methodology employed in this paper, and limitations discussed later). This constitutes 0.10% of annual US GDP over five years. We estimate that the selected programs could result in a structural impact of up to 230,000 jobs over five years (0.14% of the current labor market).

We visualize the potential upstream and downstream effects in Figure 3 by relevant NAICS code for the absolute macroeconomic impact in billions, US$ (sectoral results reported in Section 4.3). The upstream output consequences indicate industries that receive a positive economic activity impact as a result of increasing broadband demand. These are the industries that provide production impacts to the telecommunications sector to produce more infrastructure, from civil engineering (e.g., installing trenching/ducting) to electronics (active equipment, such as fiber optic, routers, etc.) to basic resource inputs (concrete, steel, etc.). In contrast, the downstream output consequences indicate industries that gain positive economic activity by utilizing broadband as a production input.

Our estimate suggests a direct impact of US $64.7 billion (0.243% of US GDP) from the three broadband investment programs and household spending, a US $45.3 billion (0.178%) implicit downstream benefit, and a US $17.2 billion (.0.68%) implicit upstream benefit.

Specifically, the model estimates that the BEAD Program may increase GDP by as much as US $84.8 billion (.381%), compared to a GDP increase for the American Connectivity Program by up to US $32.7 billion (.129%). The remaining programs could increase GDP by up to US $9.78 Bn (.038%), including the TBCP potentially contributing up to US $4 Bn (0.16%), the DLTBP up to $3.8 Bn (0.15%), and the EMMBIP up to US $2 Bn (0.08%). The confidence intervals for the ACP program in Figure 3 represent the high-end and low-end estimates for household spending.

When considering the package of all initiatives, Figure 4 visualizes the total potential indirect macroeconomic impacts, reaching as much as US $62.6 Bn. This translates to a Type I multiplier of 2.06 for the broadband investment associated with the Bipartisan Infrastructure Law. Specifically, this relates to multipliers for the BEAD program of 1.99, the ACP program of 2.30, and the additional programs of 1.95. It is additionally important to note that the multiplier associated with the Infrastructure and Jobs Act is not equal to the multiplier associated with broadband, calculated to be about 1.89 because we included induced household spending in our GDP estimates.

We can attribute ACP's high multiplier to the independent household investment that arises from the subsidy, as 82.3% is provided by the government and 17.7% is spent by households. When accounting for this, the overall Type I multiplier for the ACP and independent investment is 1.89.

Additionally, the estimated employment benefits of the broadband packages within the Bipartisan Infrastructure Law are visualized in Figure 5. The model estimates that Professional, Scientific, and Technical Services could see an employment increase of up to 29,100 (0.28% growth over five years), Manufacturing could potentially create up to 37,900 jobs (0.30% growth over five years), and service-oriented sectors are impacted considerably, with Education (25,500; 0.7% growth over five years), Information (22,200; 0.78% growth over five years), Health Care (21,600; 0.11% over five years), and Retail Trade (19,900; 0.15% over five years) seeing commensurate job growth.

These results provide insights into the economic impacts of the policy decisions associated with the BIL.

[A] GDP Effects of the BEAD Program

Augmentation of Ghosh Supply-Side Model Shock

|  | Total GDP Output |
|---|---|
| Information | $ 8.46 Bn |
| Professional, Scientific, and Technical Services | $ 7.14 Bn |
| Government | $ 3.98 Bn |
| Finance and Insurance | $ 2.55 Bn |
| Real Estate and Rental and Leasing | $ 2.3 Bn |
| Retail Trade | $ 2.29 Bn |
| Wholesale Trade | $ 2.25 Bn |
| Education | $ 2.02 Bn |
| Health Care and Social Assistance | $ 1.94 Bn |
| Administrative and Waste Management Services | $ 1.81 Bn |
| Transportation and Warehousing | $ 1.57 Bn |
| Manufacturing of Nondurable Goods | $ 1.48 Bn |
| Manufacturing of Durable Goods | $ 1.43 Bn |
| Construction | $ 1.35 Bn |
| Accomodation and Food Services | $ 1.04 Bn |
| Managment of Companies and Enterprises | $ 0.747 Bn |
| Other non-government services | $ 0.522 Bn |
| Utilities | $ 0.367 Bn |
| Mining | $ 0.338 Bn |
| Agriculture, forestry, fishing, and hunting | $ 0.233 Bn |
| Arts, Entertainment, and Recreation | $ 0.193 Bn |

0.0 2.5 5.0 7.5 10.0

Total GDP Output (US$ Billions)

[B] GDP Effects of the ACP

Augmentation of Leontief Model Shock

|  | Total GDP Output |
|---|---|
| Information | $ 4.65 Bn |
| Professional, Scientific, and Technical Services | $ 2.55 Bn |
| Manufacturing of Durable Goods | $ 1.65 Bn |
| Administrative and Waste Management Services | $ 1.18 Bn |
| Real Estate and Rental and Leasing | $ 1.17 Bn |
| Finance and Insurance | $ 0.756 Bn |
| Manufacturing of Nondurable Goods | $ 0.567 Bn |
| Wholesale Trade | $ 0.564 Bn |
| Arts, Entertainment, and Recreation | $ 0.5 Bn |
| Education | $ 0.495 Bn |
| Transportation and Warehousing | $ 0.376 Bn |
| Government | $ 0.331 Bn |
| Managment of Companies and Enterprises | $ 0.313 Bn |
| Accomodation and Food Services | $ 0.192 Bn |
| Other non-government services | $ 0.184 Bn |
| Utilities | $ 0.183 Bn |
| Construction | $ 0.137 Bn |
| Mining | $ 0.112 Bn |
| Retail Trade | $ 0.0383 Bn |
| Agriculture, forestry, fishing, and hunting | $ 0.0275 Bn |
| Health Care and Social Assistance | $ 0.00199 Bn |

0.0 2.5 5.0 7.5 10.0

Total GDP Output (US$ Billions)

[C] GDP Effects of other Programs

Augmentation of Ghosh Supply-Side Model Shock

|  | Total GDP Output |
|---|---|
| Information | $ 1.14 Bn |
| Professional, Scientific, and Technical Services | $ 0.8 Bn |
| Government | $ 0.319 Bn |
| Real Estate and Rental and Leasing | $ 0.298 Bn |
| Manufacturing of Durable Goods | $ 0.292 Bn |
| Finance and Insurance | $ 0.268 Bn |
| Administrative and Waste Management Services | $ 0.265 Bn |
| Wholesale Trade | $ 0.224 Bn |
| Education | $ 0.2 Bn |
| Manufacturing of Nondurable Goods | $ 0.171 Bn |
| Retail Trade | $ 0.166 Bn |
| Transportation and Warehousing | $ 0.154 Bn |
| Health Care and Social Assistance | $ 0.137 Bn |
| Construction | $ 0.111 Bn |
| Accomodation and Food Services | $ 0.0958 Bn |
| Managment of Companies and Enterprises | $ 0.0891 Bn |
| Arts, Entertainment, and Recreation | $ 0.0716 Bn |
| Other non-government services | $ 0.0582 Bn |
| Utilities | $ 0.0472 Bn |
| Mining | $ 0.0369 Bn |
| Agriculture, forestry, fishing, and hunting | $ 0.0197 Bn |

0 2.5 5.0 7.5 10

Total GDP Output (US$ Billions)

Type of Expenditure: Upstream, Downstream

Figure 3: Potential sectoral GDP impacts of the Bipartisan Infrastructure Law, separated by program

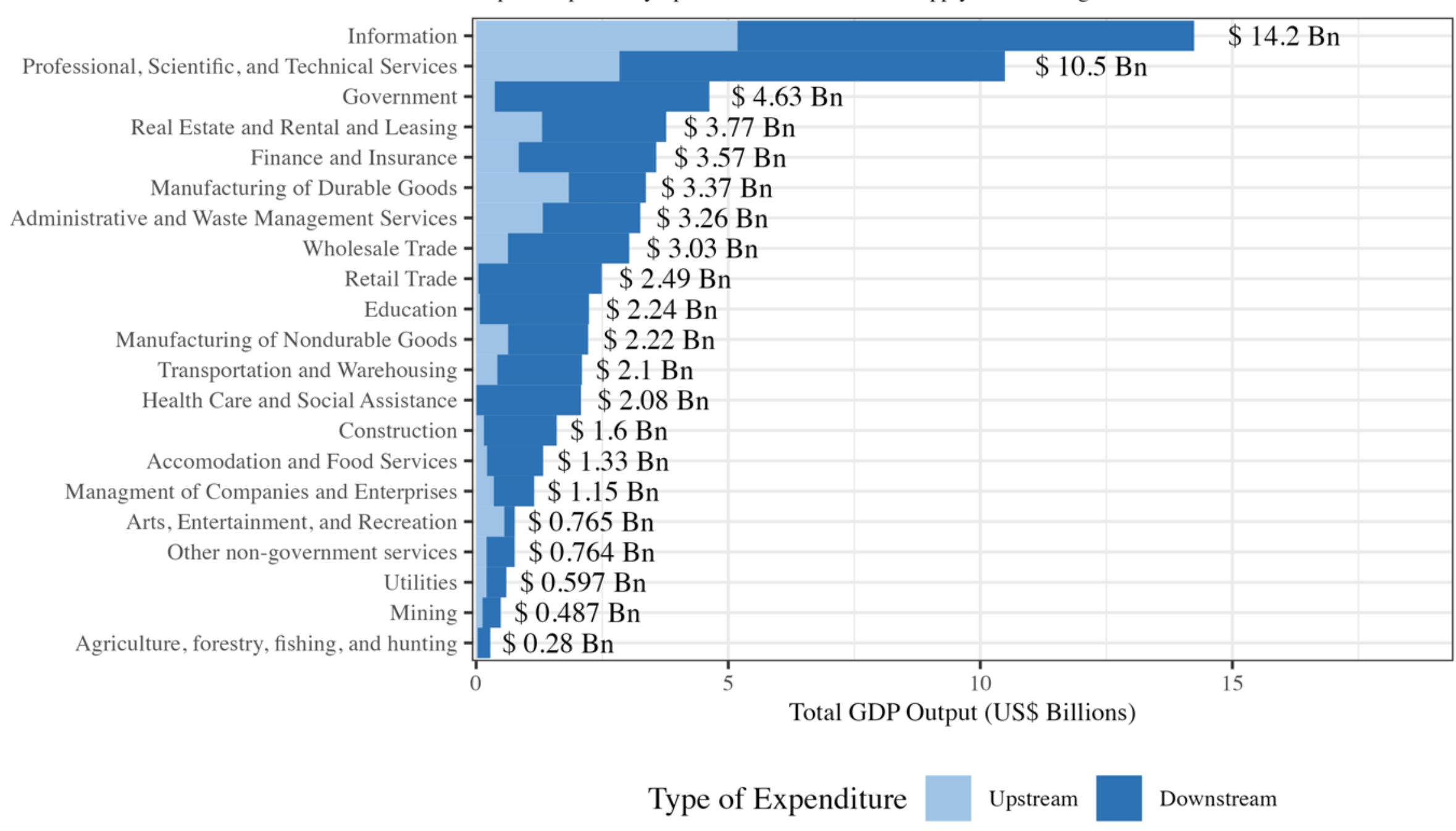

Figure 4: Total estimated sectoral output from broadband programs within the Bipartisan Infrastructure Law

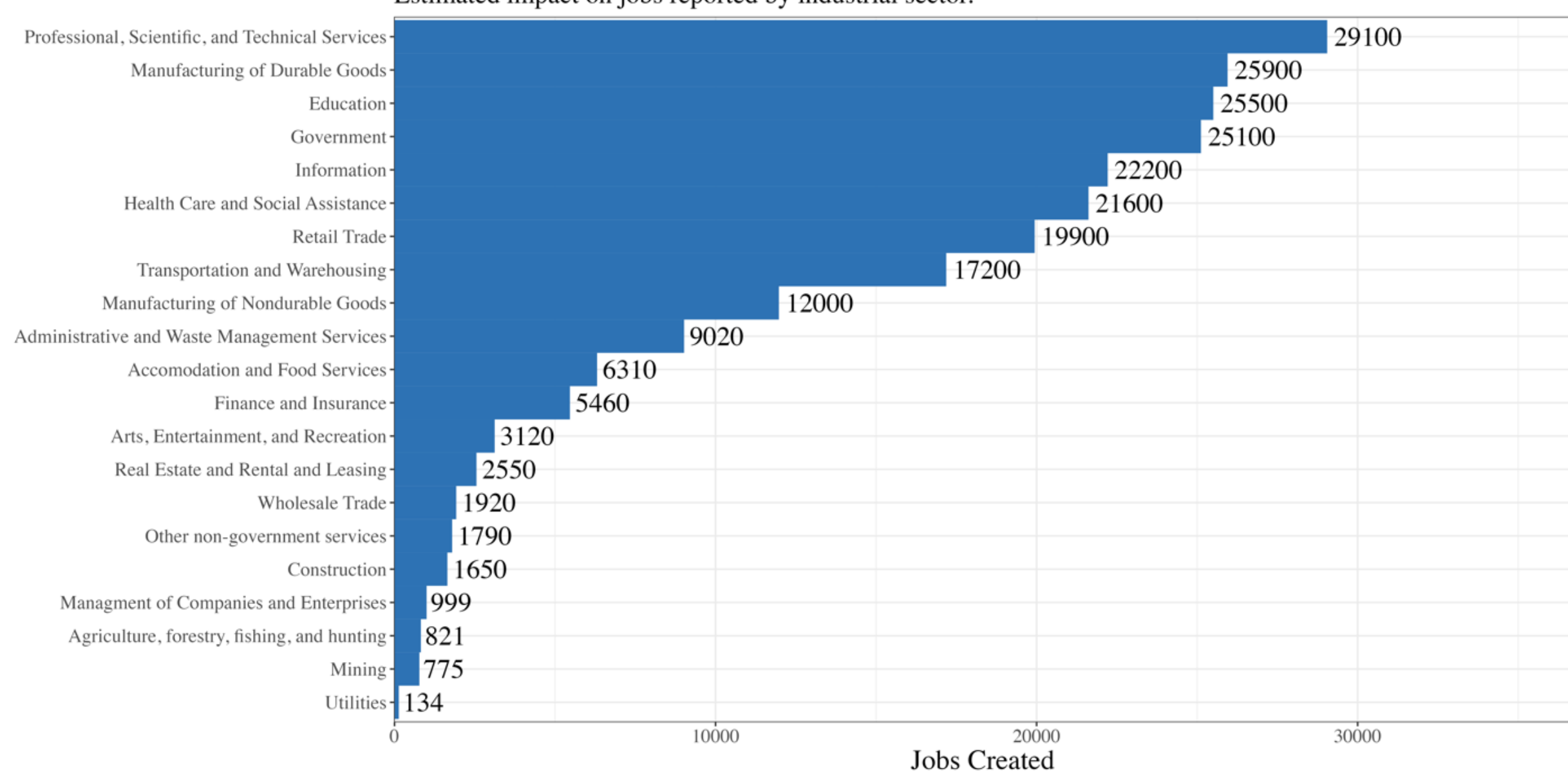

Figure 5: Total estimated sectoral employment from broadband programs within the Bipartisan Infrastructure Law

## 4.3 How are the supply chain linkages affected by allocations from the Bipartisan Infrastructure Law?

A Sankey diagram is presented in Figure 5 visualizing the key linkage impacts of these three initiatives. As illustrated, the indirect benefits ripple throughout all industries and the overall benefit of broadband infrastructure investments through the BEAD program is greater than that of subsidies in the ACP. Figure 3 displays the inter-sectoral impacts of each bill. Overall, the investment will increase Manufacturing output by US $5.59 billion, which could grow the manufacturing sector by up to .004% over the next five years.

Other than the Information sector, the Professional, Scientific, and Technical Services sector will increase by US $10.5 billion, or 0.016% over the next five years. Other sectors impacted the most are Real Estate, Rental, and Leasing; Finance and Insurance; and Administrative and Waste Management Services. The Professional, Scientific, and Technical Services sector is currently the fourth largest economic sector, behind Wholesale trade and Construction as defined by the Bureau of Economic Analysis. Those two sectors are projected to grow by US $3.03 billion (.0275%, annually) and US $1.60 billion (.017%, annually), respectively.

Sectors that are closely dependent on telecommunications are projected to grow, too. Arts, Entertainment, and Recreation, an industry that is linked to advertisement, video streaming, and other resources that depend on reliable Internet connection, are projected to grow by US $0.765 billion (.009%), Although this is one of the smallest sectors of the US economy, new connections to the Internet will spur demand for new goods and services from this sector.

Sectors that are affected the least include Agriculture, Forestry, Fishing, and Hunting ($280 Mn, .012% growth), Mining ($487 Mn, .018% growth), and Utilities ($597 Mn, .025% growth). All three industries are not closely related to the telecommunications industry, with most relying on broadband for sales, input goods, and purchase of capital goods. The relatively high growth rate speaks more to the size of the sectors in comparison to others, as these are also relatively small sectors in the US economy.

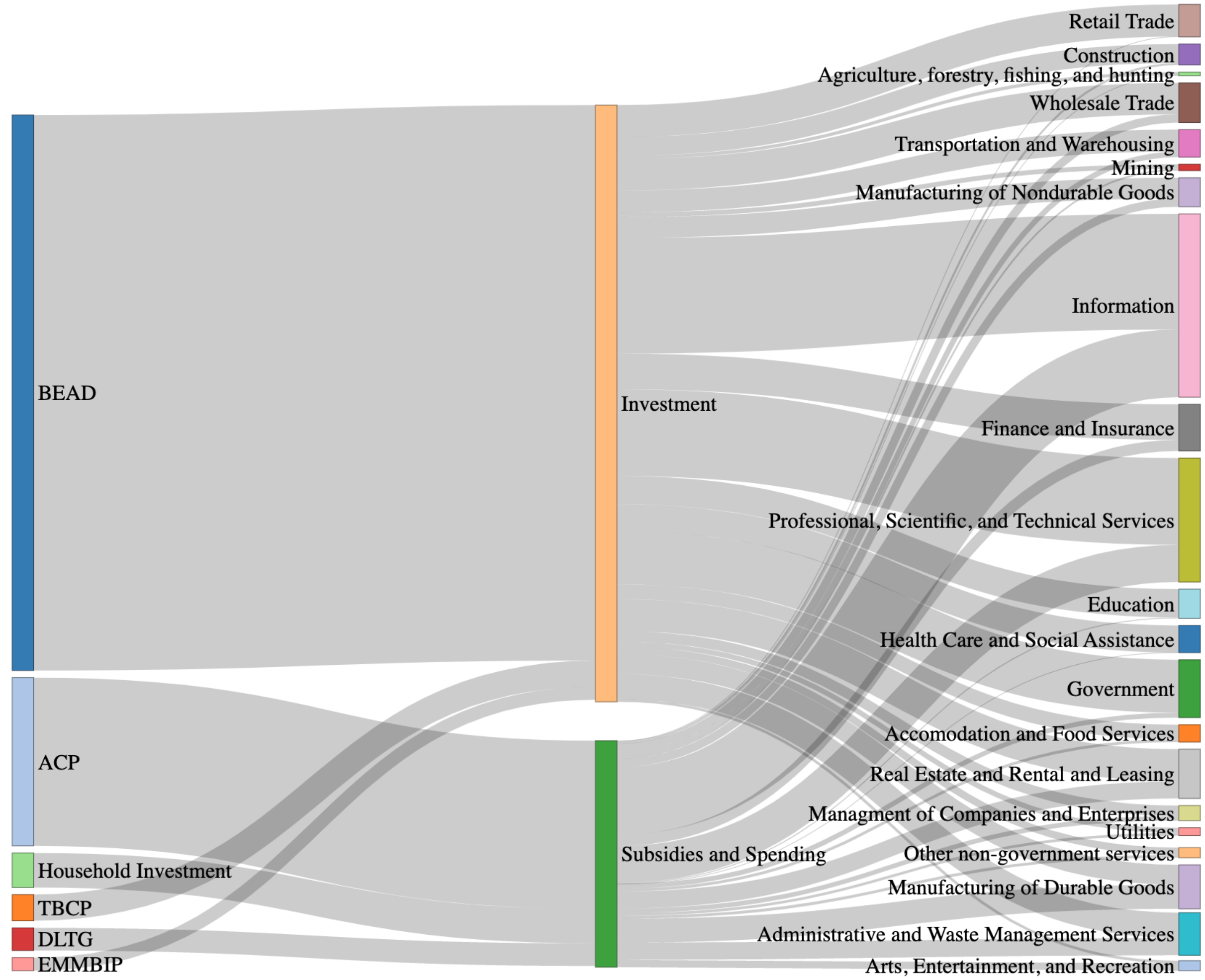

Figure 6: Sankey Diagram illustrating potential input flows from the various programs, translating to GDP impacts across a wide range of industrial sectors.

# 5 Discussion

In this Section 5, a discussion will be undertaken where the previously articulated results are evaluated within the broader context of the research questions.

## 5.1 To what extent does the Bipartisan Infrastructure Law allocate funding to unconnected communities in need?

Generally, most states are organized inversely between BEAD Allocation and the supply of existing broadband infrastructure. This result indicates that the broadband investment programs in the Bipartisan Infrastructure Law allocate funding effectively, targeting communities without broadband supply or those with a significant price barrier. Indeed, our analysis finds that regions with poorer broadband infrastructure receive commensurate broadband investment via the programs available, generally suggesting the allocated capital appears to be directed where there is a clear need (as opposed to politically motivated capital allocation). For example, states with lower-than-average ACP enrollment see a greater BEAD allocation, and this may be explained by the fact that households in poorly connected areas have less reason to enroll in a substandard service. Equally, areas with higher-than-average ACP enrollment see fewer BEAD resources allocated.

While this pattern is consistent, there is one extreme outlier whereby the BEAD program intends to spend almost US $40k to connect each Alaskan household to broadband. Alaska adheres to the general trend we established in Section 4.1, whereby only about 7.2% of Alaskan homes have enrolled in the ACP, which could suggest fewer households have broadband access. Due to the low population density in Alaska, deploying broadband infrastructure will take significantly more time and investment (Espín & Rojas, 2024). Instead of connecting to existing broadband networks within the continental United States, Alaskan providers often need to lay large quantities of greenfield infrastructure over long distances within the Alaskan wilderness. To put this investment into perspective, US $38,802 is more than half of the median household income (56%) in the United States (US $69,600) (Guzman, 2022). Connecting Alaskan households to fixed broadband rather than exploring satellite broadband access could prove a costly policy choice. An Alaskan community with ten unconnected homes could receive an average of roughly US $390,000 to connect to wireline broadband, broadly equating with hiring a teacher for five years (US Bureau of Labor Statistics, 2022) or establishing scholarships for about 50 students to attend the University of Alaska (University of Alaska, 2024).

Another outlying case is Texas, where a significant rural population lowers its per-capita allocation relative to other rural states. Even though Texas will receive the largest grant by far, each unconnected household will only see about US $2-3k to connect to broadband. This per-capita allocation is lower than average, and the ACP connectivity (13.8% enrollment) is approximately average. The discrepancy could result from the high number of unconnected households in Texas - which means the high allocation disperses over a greater number of houses than in other states. Additionally, the spatial distribution of homes in Texas is likely to be denser than in Alaska, and most homes will be closer to an urban area or major highway with substantial local or long-distance fiber infrastructure to connect to. These conditions make it less costly to connect Texan households to the state's broadband network.

Some states, like Ohio and New York, have high ACP enrollment and low BEAD allocations, indicating that a significant barrier to broadband adoption is potentially from lower-income households being unable to afford the service (rather than a lack of existing infrastructure). ACP enrollment in Louisiana is encouraging and shows considerably higher registration than in other states. For example, over 26% of households are enrolled in the program, compared to the nationwide average of 13.3%. Similarly, Ohio (20.7%), Kentucky (22.6%), and New Mexico (20.2%) stand out as states that have succeeded in efforts to ensure that their communities connect to the ACP. States with very high enrollment levels should be concerned as to whether this program may secure longer-term funding into the future. State and local governments should actively develop policies which encourage lower-income households to sign-up to federally-funded programs, with ACP being a recent example.

## 5.2 What are the GDP and employment impacts of the three funding programs within the Bipartisan Infrastructure Law?

We find that the US $61.7 billion investment from the selected programs of the Bipartisan Infrastructure Law could lead to a total potential macroeconomic impact of up to $127.3 billion, given an indirect impact of $62.6 billion (49.2%). This is based on the direct investment being spent on building out (predominantly) new fixed broadband infrastructure for cost items such as civil engineering, active equipment deployment, and the associated labor necessary to connect households to high-speed broadband Internet. Whereas, the indirect investment represents both industries utilizing broadband to grow their customer base and the multiplier effects of employed labor undertaking secondary spending in the economy.

Our research estimates that overall, the Bipartisan Infrastructure Law could have a return multiplier as high as 2.06, and broadband investment could have a multiplier as high as 1.89. The multiplier of the investment package is higher than the calculated Type I multiplier for broadband investment because the former included induced spending stemming from household investment. These results match other infrastructure investment estimates in literature, such as Prieger (2020) which estimated a multiplier of 1.61-1.76; Rohman & Bohlin (2012), which estimated an output multiplier of 1.43-3.6 for the US, Germany, and the UK; Sosa & Van Audenrode (2011) which cite a private investment multiplier of 3.08; and Lobo et al. (2008), which found a county income multiplier of 1.99. Therefore, this spending could boost GDP growth while potentially providing long-term opportunities for (particularly rural) households at a time when our reliance on high-speed broadband connectivity is growing. Thus, there is a need to overcome market failure in areas of current poor broadband service. Proponents argue that the lack of infrastructure necessitates the passage of the BIL, as the return on investment to national GDP outweighs the cost to the taxpayers. Additionally, as the estimated $127.3 billion in GDP trickles through the macroeconomy, there may be other economic benefits, ranging from productivity gains to increased entrepreneurship. Many researchers have noted that broadband has significant non-economic benefits, such as sustainability and quality of life improvements that cannot be measured using GDP (Valentín-Sívico et al., 2023; Van Parys & Brown, 2023).

This research also estimates that the broadband investment programs within the Bipartisan Infrastructure Law could potentially create up to 230,000 jobs with the Professional, Scientific, and Technical Services Sector, Manufacturing, Education, Information, and Health Care being impacted considerably. These results align with other reported job growth estimations in the literature. One study included that the growth of broadband-related jobs is primarily within service-related industries. This matches our estimation, as four of the top five impacted sectors are primarily service-focused (Katz & Suter, 2009).

The broadband programs within the Bipartisan Infrastructure Law are the largest federally-directed investment in broadband infrastructure to date. The program's goals are similar to those of the Rural Electrification Act (REA) in the 1930s, established by Franklin D. Roosevelt as an investment associated with the New Deal (*Record Group 221*, 1934). At the time, many detractors argued that the program was too expensive. However, as evidence has shown, the investment laid the foundation for nearly a century of economic development, improving the quality of life of citizens and expanding business opportunities, especially for rural regions (Bouzarovski et al., 2023; Olanrele, 2020; Tierney, 2011).

Similarly, many critics of the programs take issue with the significant government investment at a time when fiscal headroom is at a minimum. Recently, many congressional representatives have voiced their concerns about government spending, stating "...we have no money" and "there's no money in the house" (Davis, 2023; Sforza, 2023). Concern has been growing about the current administration financing government spending through deficit spending, for example, with the US budget deficit effectively doubling in 2023, with the ongoing expansion of a range of federal programs, such as BEAD, ACP, and the TBCP (Gopinath, 2023; Rappeport & Tankersley, 2023).

Some valuable policy implications of these findings are that the multiplier effects might not necessarily outpace maturity for treasury bonds. The federal government relied on deficit spending to fund these programs, which means the United States Department of the Treasury borrowed money by issuing bonds. An analysis of bond maturity suggests that $1 in debt at a roughly 4% coupon rate with a 2% yearly inflation rate would cost the US government $1.22-2.21 to borrow, depending on how often the interest accrues. Since $1 of debt generates an estimated roughly US $1.89 to US $2.06 dollars in GDP, and including the fact that IO modeling can overestimate growth, each dollar invested into these programs may cost more than it generates in the short-run. However, as proponents have argued, these programs focus on long-term quality of life improvements, development, and entrepreneurship, and short-term economic loss could be a worthwhile trade-off to try to close the digital divide. More research needs to be undertaken to understand this cost-benefit trade-off.

Additionally, bond maturity analysis does not include the employment and induced spending effects of the broadband investment packages within the Bipartisan Infrastructure Law. As proponents have argued, investments in passive Internet infrastructure have been shown to increase job availability and subsequently improve quality of life in underserved areas. While these investments may not outpace bond maturity in the short-run, the job creation and induced spending could potentially mean these investments will see a higher return in the long-run.

Finally, an important policy implication arises from comparing multipliers between different sectors of the economy. A $1 investment into broadband could structurally produce up to $1.89 for the economy. However, similar structural investments into Social Assistance, Wholesale Trade, and Education would produce potentially $1.81, $1.74, and $1.64 for the economy. This implies that, in the short-run, investments into broadband infrastructure could generate more GDP growth than investments into these industries.

## 5.3 How are supply chain linkages affected by allocations from the Bipartisan Infrastructure Law?

Undertaking a key linkage analysis of the industrial sectors most affected from this investment, we quantify the impacts on both upstream and downstream industries (depending on whether the investment is targeted at the supply-side or demand-side). The information sector (NAICS 51) is expected to indirectly grow by as much as US $14.2 billion, which is greater compared to all other industries. This growth is distinguished by up to $9.06 billion (63.6%) in downstream impact and $5.18 billion (36.4%) in upstream impact. When including the direct impact, the Bipartisan Infrastructure Law investment would see this sector increase by over 0.15% over the next five years, due to greater reliance on Internet services, thanks to expanded access to wireline broadband.

The sector with the second-largest growth is estimated to be the Professional, Scientific, and Technical Services sector (NAICS 54). Growing by as much as $10.5 billion, through $7.65 billion downstream (72.9%) and $2.84 billion upstream (27.1%). As businesses and economies gain access to wireline broadband and high-speed Internet, they can use these services to organize and communicate, contributing to the upstream effects. The large downstream quantity is likely because of broadband's dependency on this sector for development and service.

Interestingly, in sectors such as Retail Trade (NAICS 44-45, $2.49 billion) and Health Care and Social Assistance (NAICS 62, $2.08 billion), downstream effects ($2.45 billion, 98.4% and $2.08 billion, 99.8%, respectively) outweigh the upstream effects ($.043 billion, 1.6% and $.0022 billion, 0.163%, respectively), possibly because these industries are more reliant on broadband as a production input. Even though post-COVID trends suggest consumers are generally returning to physical brick-and-mortar stores, more consumers now value the access and convenience of online retail and purchasing, so the retail industry is becoming increasingly reliant on broadband connectivity to make sales (Brüggemann & Olbrich, 2023; Inoue & Todo, 2023; Sen et al., 2023). Likewise, Health Care is also becoming more reliant on broadband for record storage and interhospital communication. The growth in these sectors implies that they stand to gain significantly more from an increased availability of broadband rather than the increased demand for broadband.

The opposite is true for the Arts, Entertainment, and Recreation Services ($.765 Bn NAICS 71), a sector that is affected more by the upstream effect ($0.558 billion, 72.9%) rather than the downstream ($0.207 billion, 27.1%). As consumers connect to new or faster broadband, they also connect to online content sources. However, instead of the Entertainment industry relying on broadband, the consumer utilizes these services to access content. Therefore, it is reflected in the demand-side of the model rather than the supply-side, as Entertainment benefits more from the increased demand for broadband rather than from its increased availability. A similar rationale is likely the cause for the opposite split in the Agriculture, Forestry, Fishing, and Hunting sector (NAICS 11, $.280 billion), which is also the sector that is projected to grow the least, with the downstream impact ($0.250 billion, 89.3%) far outweighing the upstream impact ($0.031 billion, 11.7%).

In general, the results we find here suggest that the most interconnected industries are projected to see the largest increase in growth from the investment programs, such as the Information sector (NAICS 51, $14.2 billion) and Professional, Scientific, and Technical Services (NAICS 54, $10.5 billion).

# 6 Conclusion

In this paper, we assessed the potential structural macroeconomic impacts of the Broadband, Equity, Access and Deployment program, American Connectivity Program, and Tribal Broadband Connectivity Program. We firstly evaluated the program allocation and enrollment on a state-by-state basis, Secondly, we estimated the GDP impact by using national accounting data to report the downstream impacts via the Ghosh Supply-side model and upstream impacts via the Leontief Demand-side model. To our knowledge, this is the first structural macroeconomic assessment of the broadband infrastructure programs within the Bipartisan Infrastructure Law.

In general, we found that higher BEAD allocations target areas with poor current broadband supply. While this may be logical, as it illustrates funding going to areas most in need, this could not have been assumed *a priori* given politically-motivated funding is not always rationally allocated. We also found some outliers with BEAD allocations, for example, with Alaska receiving roughly US $39,000 per unconnected household. Moreover, Louisiana ACP enrollment is twice as high as the mean state enrollment, with around 26.6% of homes enrolling (versus a mean of 13.3%).

In terms of macroeconomic impact, the total direct contribution to US GDP by the program could be as high as US $84.8 billion, $32.7 billion, and $9.78 billion for the BEAD program, ACP, and additional programs, respectively. Overall, the broadband allocations could expand US GDP by $127.3 billion, with a multiplier of 1.97 (with a calculated Type I Multiplier of 1.89). The NAICS industries that may benefit the most from these programs include the Information sector (NAICS 51, $14.2 billion), the Professional, Scientific, and Technical Services Sector (NAICS 54, $10.5 billion), and the Manufacturing Sector (NAICS 31-33, $5.59 billion). Insights about the supply-chain industry were also evaluated, including the relative interdependencies of sectors on broadband telecommunications and how certain sectors were more reliant on high-speed, reliable Internet as a value-added input good than other sectors (Agriculture, Forestry, Fishing, and Farming and Hunting; NAICS 11; 89.3% downstream and Retail Trade; NAICS 44-45; 99.8% downstream).

In terms of employment impact, the total job growth could be as high as 230,000. Service-oriented industries like the Professional, Scientific, and Technical Services Sector; Education; Information; and Health Care and Social Assistance could see growth of up to .28%, 0.7%, 0.78%, 0.11% over the next five years. As a result of investment into wireline broadband deployment, the Manufacturing sector could see the creation of up to 37,900 jobs (0.30% growth over the next five years). The least impacted sectors include Agriculture, Forestry, Fishing, and Hunting; Mining; and Utilities.

As with any method, there are relevant limitations which are worth highlighting. Firstly, the IO approach in this paper is static and does not account for potential dynamic impacts in the macroeconomy as supply and demand forces play out in a market environment. Thus, we do not account for structural changes or how measured stimulus would alter sectoral interdependencies. Like other IO approaches, we assume constant proportionality of industry inputs, so there are no changes to the original technical coefficients matrix regardless of sectoral augmentation from additional output, increased productivity, or novel technologies (Galbusera & Giannopoulos, 2018; Miernyk, 1966). Moreover, this approach depends on constant returns to scale, with no decreasing marginal cost of production as output increases. Importantly, this IO model approach produces the upper bound on GDP impacts, so growth could be lower than projected (Jones, 1968; Oosterhaven, 1988; Walheer, 2020). Additionally, double-counting may have affected these model estimates due to counting the increase of one sector without recognizing the subsequent resulting loss of another. Because this modeling architecture holds these losses exogenous, they are not reflected in the end result. Likewise, estimating household spending leaves out shifts due to budget constraints - without more data, IO models hold these exogenously. To balance the effect of double-counting, value-added and demand multiplier effects were considered independently and did not overlap in the presented results.

Future research could expand the methodology in this paper to focus on dynamic macroeconomic modeling strategies to study how spending increases from building new broadband infrastructure compares to the estimates produced here. There are also future opportunities within the realm of static modeling, as future research could analyze how carbon emissions, gross capital formation, or other economic markers are affected by these policies. Moreover, future research within static modeling could focus on creating a tailor-made broadband column for Input-Output analysis to provide enhanced structural insight into potential supply chain impacts. Future research could also focus on analyzing disaggregated data, including the impact of socioeconomic status and other human factors on broadband availability and access, to make conclusions about availability and accessibility on a more precise scale than the state-level. Additionally, future studies could examine how to better allocate funding towards future infrastructure investments, ideally maximizing the effectiveness of allocations through the BEAD and ACP enrollment programs.

# 7 Acknowledgements

We thank Tony To for reviewing an earlier version of this manuscript and providing comments. We would also like to thank two anonymous reviewers who provided feedback which significantly improved the manuscript.

# 8 Appendix

| Sector | NAICS | Sections Referenced |
| --- | --- | --- |
| Agriculture, Forestry, Fishing, and Hunting | 11 | 4.3; 5.3; 6 |
| Mining | 21 | 4.3 |
| Utilities | 22 | 4.3 |
| Construction | 23 | 2.2; 4.3 |
| Manufacturing | 31-33 | 2.1; 4.3 |
| Wholesale Trade | 42 | 4.3 |
| Retail Trade | 44-45 | 5.3 |
| Transportation and Warehousing | 48-49 | 2.1; 2.2 |
| Information | 51 | 2.1; 2.2; 4.2; 4.3; 5.2; 5.3 |
| Finance and Insurance | 52 | 4.3 |
| Real Estate and Rental and Leasing | 53 | 4.3 |
| Professional, Scientific, and Technical Services | 54 | 4.3; 5.3 |
| Management of Companies and Enterprises | 55 | - |
| Administrative and Waste Management Services | 56 | 4.3 |
| Education | 61 | 1; 2.1; 2.3; 3.3 |
| Health Care and Social Assistance | 62 | 5.3 |
| Arts, Entertainment, and Recreation | 71 | 4.3; 5.3 |
| Accommodation and Food Services | 72 | - |
| Other non-government services | 81 | - |
| Government | 92 | 3.1 |

Supplementary Information Table 1: NAICS classifications for referenced sectors
(US Census Bureau, 2023)